\documentclass[preprint]{aastex}
\usepackage{emulateapj5}
\usepackage{apjfonts}
\usepackage{natbib}

\input psfig.tex

\def\aa{{A\&A}}
\def\aas{{ A\&AS}}
\def\aj{{AJ}}
\def\al{$\alpha$}
\def\bet{$\beta$}
\def\amin{$^\prime$}
\def\annrev{{ARA\&A}}
\def\apj{{ApJ}}
\def\apjs{{ApJS}}
\def\asec{$^{\prime\prime}$}

\def\deg{$^{\circ}$}

\def\e#1{$\times$10$^{#1}$}
\def\etal{{et al. }}

\def\hst{{\it HST}}

\def\lamb{$\lambda$}
\def\lax{{$\mathrel{\hbox{\rlap{\hbox{\lower4pt\hbox{$\sim$}}}\hbox{$<$}}}$}}
\def\gax{{$\mathrel{\hbox{\rlap{\hbox{\lower4pt\hbox{$\sim$}}}\hbox{$>$}}}$}}
\def\simlt{\lower.5ex\hbox{$\; \buildrel < \over \sim \;$}}
\def\simgt{\lower.5ex\hbox{$\; \buildrel > \over \sim \;$}}
\def\lum{erg s$^{-1}$}
\def\mbh{{$M_{\rm BH}$}}

\def\mnras{{MNRAS}}
\def\nat{{Nature}}
\def\pasp{{PASP}}

\def\percm2{cm$^{-2}$}

\def\solmass{$M_\odot$}

\slugcomment{To appear in The Astrophysical Journal.}
\shorttitle{RADIO EMISSION AND BLACK HOLE MASS}
\shortauthors{HO}

\begin{document}

\title{On the Relationship Between Radio Emission and Black Hole Mass in 
Galactic Nuclei}

\author{Luis C. Ho}
\affil{The Observatories of the Carnegie Institution of Washington, \\
813 Santa Barbara St., Pasadena, CA 91101}

\begin{abstract}
We use a comprehensive database of black hole masses (\mbh) and nuclear 
luminosities to investigate the relationship between radio emission and \mbh.
Our sample covers a wide range of nuclear activity, from nearby inactive 
nuclei to classical Seyfert~1 nuclei and luminous quasars.  Contrary to some
previous studies, we find that the radio continuum power, either integrated 
for the entire galaxy or isolated for the core, correlates poorly with \mbh.
The degree of nuclear radio loudness, parameterized by the radio-to-optical 
luminosity ratio $R$, also shows no clear dependence on \mbh.  Radio-loud 
nuclei exist in galaxies with a wide range of \mbh, from $\sim 10^6$ \solmass\ 
to a few\e{9} \solmass, and in a variety of hosts, from disk-dominated spirals 
to giant ellipticals.  We demonstrate that $R$ is strongly inversely 
correlated with $L/L_{\rm E}$, the ratio of nuclear luminosity to the 
Eddington luminosity, and hence with mass accretion rate.  Most or all of the 
weakly active nuclei in nearby galaxies are radio-loud, highly sub-Eddington 
systems that are plausibly experiencing advection-dominated accretion.
\end{abstract}


\section{Introduction}

Efforts to search for massive black holes (BHs) in the centers of nearby 
galaxies have made rapid progress in the last few years (Kormendy \& 
Richstone 1995; Richstone et al. 1998; van~der~Marel 1999; Ho 1999a; Kormendy 
\& Gebhardt 2001).  Kinematic observations of a significant number of galaxies 
have yielded evidence for central dark objects with masses $\sim 10^6-10^9$ 
\solmass, which can be plausibly interpreted as massive BHs.  In two cases, 
namely the center of the Milky Way and NGC 4258, the extraordinarily high 
density of the dark matter appears to rule out all reasonable astrophysical 
alternatives to a single collapsed object (Maoz 1998).  To date, BH masses 
are available for $\sim$40 inactive or weakly active galaxies from 
observations of spatially resolved kinematics and for a comparable number 
of bright active galactic nuclei (AGNs) from ``reverberation mapping'' of their 
broad-line regions (see \S\ 2.1 and Table~1), sufficient to encourage 
preliminary examinations of statistical relationships between BH masses and 
global properties of the galaxies.  Two correlations have emerged for the 
weakly active galaxies: (1) \mbh\ correlates with $L_{{\rm bulge}}$, the 
optical luminosity ($\propto$ mass) of the bulge component of the galaxy 
(Kormendy 1993; Kormendy \& Richstone 1995; Magorrian et al. 1998; Ho 1999a; 
Kormendy \& Gebhardt 2001), and (2) \mbh\ correlates even more strongly with 
$\sigma_e$, the luminosity-weighted stellar velocity dispersion within the 
bulge effective radius $r_e$ (Gebhardt et al. 2000b; Ferrarese \& Merritt 
2000).  By contrast, \mbh\ depends little on disk properties (Kormendy et al. 
2001).  These empirical relations hold great promise for furthering our 
understanding of the formation of massive BHs, the formation of galaxy 
spheroids, and the apparent close connection between the two.

Another correlation which has received considerable attention is that between 
\mbh\ and radio emission.  Franceschini, Vercellone, \& Fabian (1998) compiled 
a sample of the dozen BH masses known as of mid-1997 and showed that they 
evidently scale with radio luminosity, approximately of the form $L_{\rm rad}\,
\propto\, M_{\rm BH}^{2.5}$, where $L_{\rm rad}$ is measured at 5~GHz (6~cm).  
Surprisingly, the correlation is as tight, if not tighter, for the total radio 
emission integrated over galactic scales than it is for the core emission 
alone.  They (see also Di~Matteo, Carilli, \& Fabian 2001) argue that the 
functional form of the $L_{\rm rad}-M_{\rm BH}$ relation arises naturally if 
the BH accretion takes the form of an advection-dominated accretion flow (ADAF; 
see Narayan, Mahadevan, \& Quataert 1998b and Quataert 2001 for reviews), 
fueled by hot gas from a large-scale spherical inflow.  Moreover, the steepness 
of the relation gives it considerable leverage in predicting $M_{\rm BH}$ 
efficiently from $L_{\rm rad}$, a readily available quantity.  Several 
subsequent studies similarly considered related samples of nearby galaxies, but 
their results have been mixed (Yi \& Bough 1999; Salucci et al. 1999; Laor 
2000; Di~Matteo et al. 2001).  Discussion of the apparent dependence of \mbh\ 
on radio luminosity and the radio-loudness parameter ($R$) recently has also 
surfaced in the context of more active galaxies such as Seyferts and quasars 
(McLure et al. 1999; Nelson 2000; Laor 2000; McLure \& Dunlop 2001; 
Lacy et al.  2001; Gu, Cao, \& Jiang 2001).
 
This paper reexamines the relationship between radio emission and \mbh\ in 
light of the most up-to-date samples of active and weakly active galaxies with 
reliable BH mass measurements.  We show that the loose trend between 
integrated radio luminosity and \mbh\ is largely indirect, a 
consequence of more fundamental correlations between radio luminosity and 
bulge mass on the one hand, and between bulge mass and \mbh\ on the other.  
The distribution of nuclear radio luminosity as a function of \mbh\ is more 
physically grounded, but its large scatter renders it ineffective as a 
statistical tool to predict \mbh.  The $R$--\mbh\ relation disappears 
altogether when one considers AGNs with a broad range of intrinsic 
luminosities.  However, we find that $R$ is strongly related to the mass 
accretion rate.

\section{The Database}

We begin with a fairly detailed documentation of the data used in the 
subsequent analysis.  Particular attention is paid to the source of 
the \mbh\ measurements (Table~1) and photometric parameters (Table~2).  

\clearpage
\begin{figure*}[t]
\centerline{\psfig{file=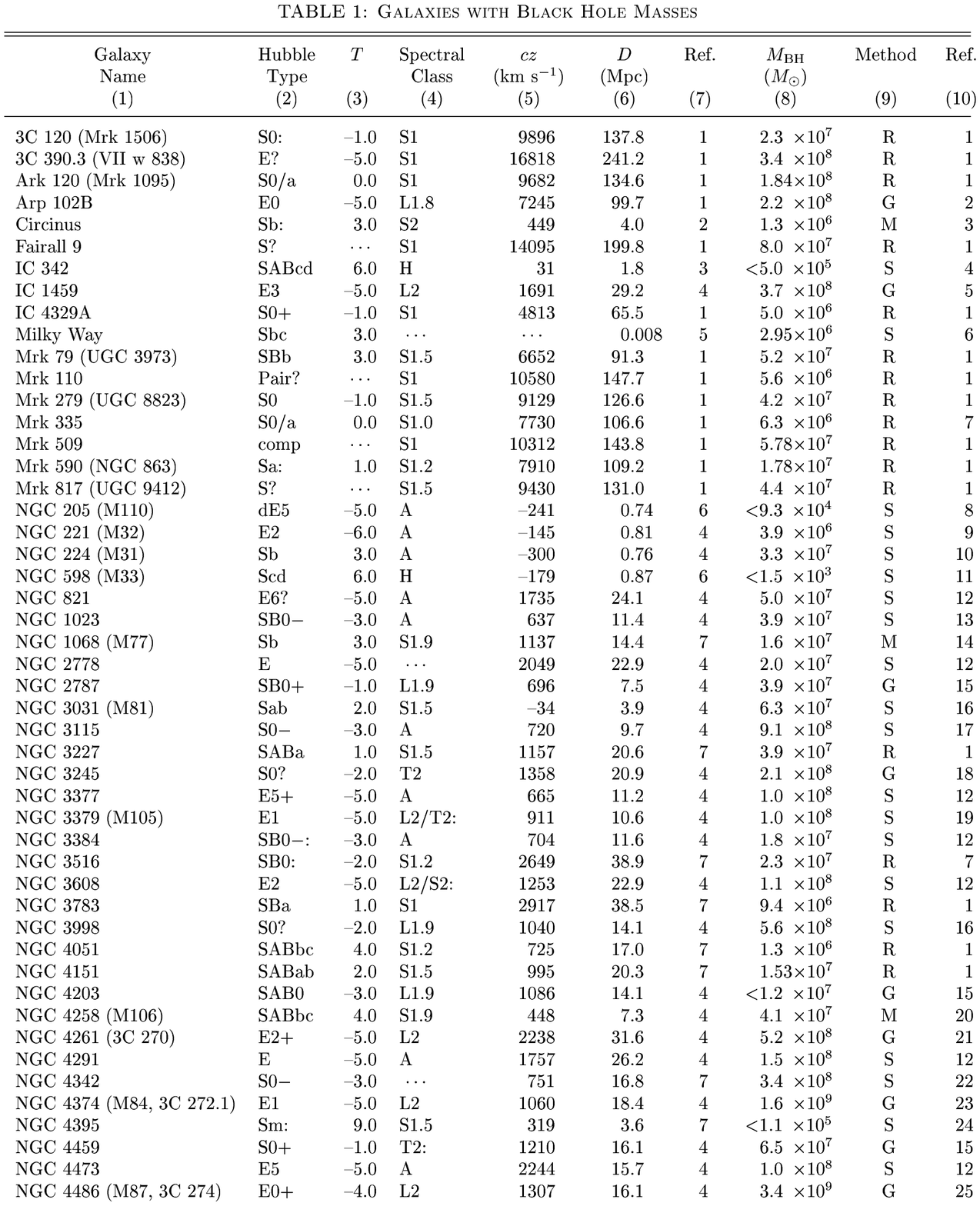,width=18.5cm,angle=0}}
\end{figure*}

\clearpage

\begin{figure*}[t]
\centerline{\psfig{file=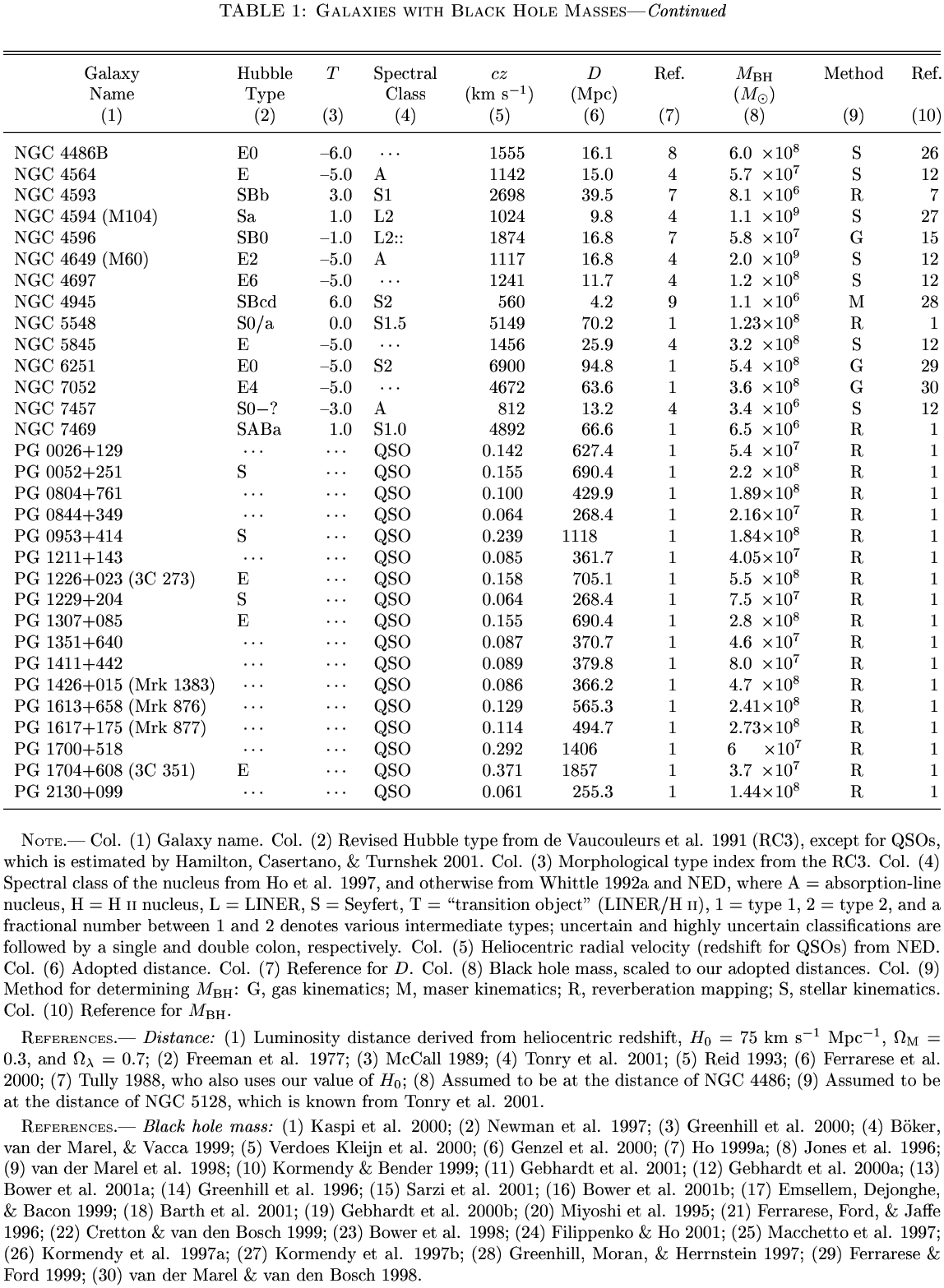,width=18.5cm,angle=0}}
\end{figure*}

\clearpage
\begin{figure*}[t]
\centerline{\psfig{file=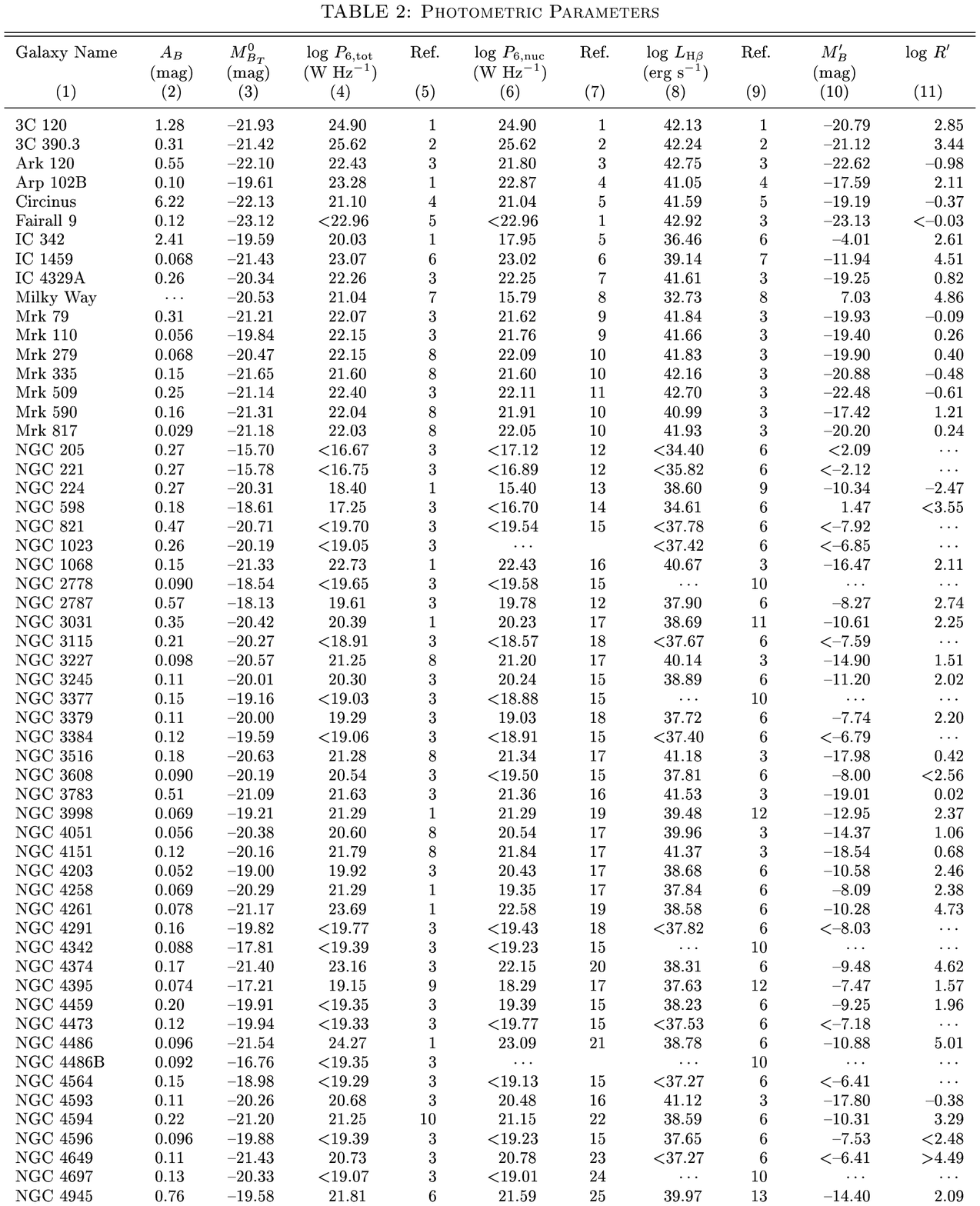,width=18.5cm,angle=0}}
\end{figure*}
 
\clearpage
\begin{figure*}[t]
\centerline{\psfig{file=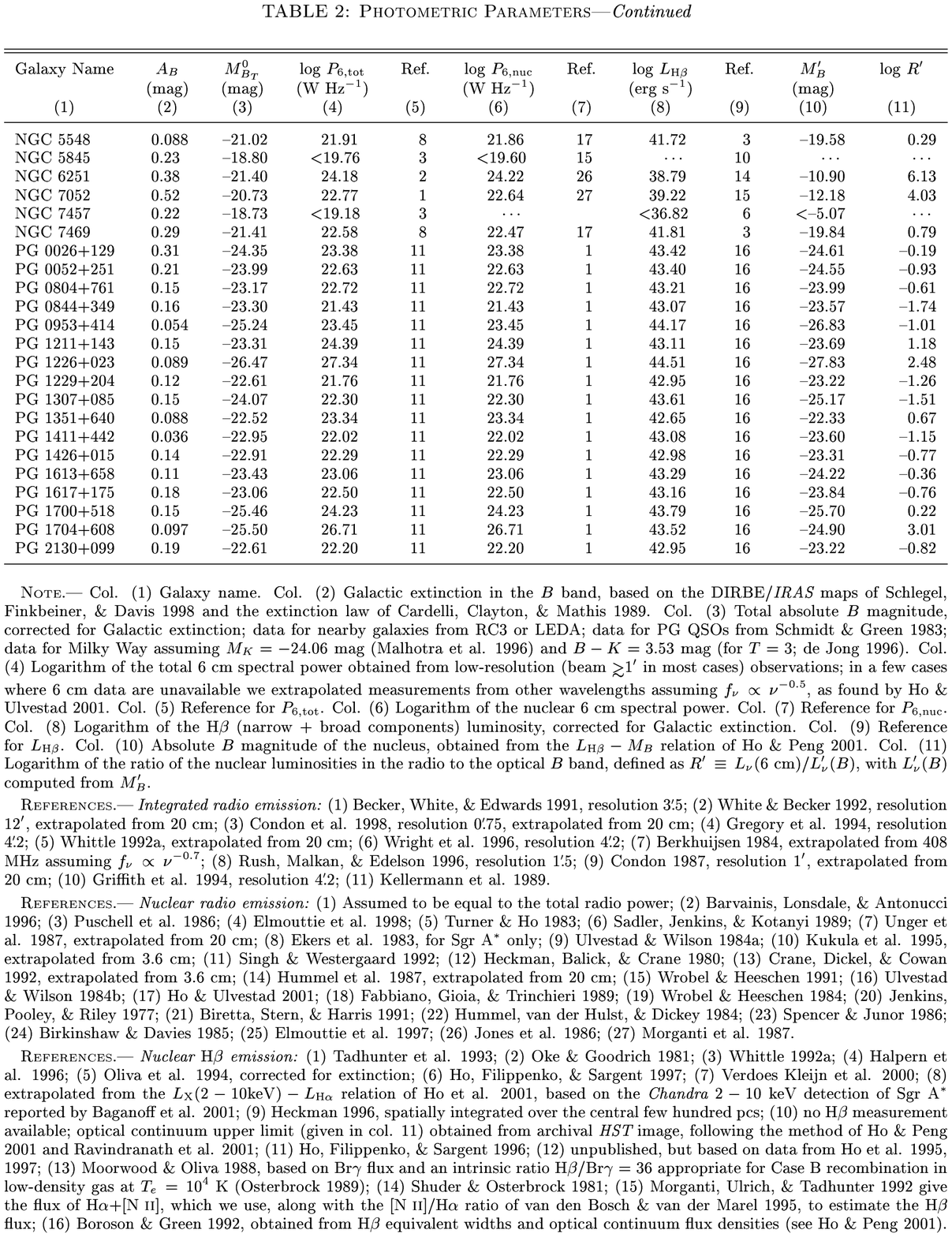,width=18.5cm,angle=0}}
\end{figure*}
\clearpage

\subsection{Black Hole Masses}

Our primary sample of BH masses comes from spatially resolved observations
of gas and stellar kinematics.  Reviews of
these methods can be found
in Kormendy \& Richstone (1995) and Ho (1999a).  Table~1 represents an update
of the compilation of BH masses given in Ho (1999a), supplemented with
newly published values. The majority of the gas measurements consist of 
optical observations of ionized gaseous disks, all done with the {\it Hubble 
Space Telescope} (\hst); four galaxies (Circinus, NGC 1068, NGC 4258, and 
NGC 4945) exploit the availability of strong water maser emission to probe the 
nuclear kinematics using radio interferometry. The broad-line radio galaxy 
Arp 102B presents a special case in which a periodic signature in its 
emission-line light curve has been interpreted as the orbital period of an 
accretion disk (Newman et al. 1997).  With the exception of Sgr A$^*$ in the 
Galactic Center, whose mass has been determined through proper motions and 
radial velocities of individual stars (Eckart \& Genzel 1997; Ghez et al. 
1998; Genzel et al.  2000), all the stellar-based masses come from integrated 
spectroscopy, a few from the ground, but largely from \hst.

Magorrian et al. (1998) published \mbh\ for a significant number of nearby 
early-type galaxies based on axisymmetric, two-integral dynamical modeling of 
\hst\ images and ground-based stellar spectroscopy.  We do not use these masses,
however, because they are subject to considerable uncertainties due to the 
simplified modeling and the low spatial resolution of the spectra.  As 
commented by a number of authors (van~der~Marel 1999; Ferrarese \& Merritt 
2000; Gebhardt et al. 2000b), the two-integral assumption will tend to 
overestimate \mbh.  We illustrate this explicitly for the 16 galaxies 
in Magorrian et al.'s sample that overlap with the sample assembled in 
Table~1 (Fig.~1).   The average offset for these objects is a factor of 3.3.
Figure~1{\it b}\ indicates that the discrepancy seems to be slightly more
severe for ``core'' galaxies than for ``power-law'' galaxies, systems with and 
without a resolved break in their inner surface brightness profiles, 
respectively (Lauer et al. 1995; Faber et al. 1997).  As a class, core 
galaxies tend to be luminous, pressure-supported systems with boxy isophotes 
(e.g., Faber et al. 1997; Ravindranath et al. 2001), precisely those most 
prone to having an anisotropic velocity distribution and thus most ill-suited 
for two-integral modeling.  Figure~1{\it b}\ offers mild support for this
picture.

We include several meaningful upper limits.  Four come from stellar-kinematical 
constraints placed on the central nuclear star cluster (IC~342, M33, NGC 
205, and NGC 4395), and a fifth derives from \hst\ gas kinematics (NGC 4203).

Direct dynamical measurements become unfeasible for luminous or distant AGNs.  
The bright continuum emission of the active nucleus nearly always overpowers 
the stellar absorption lines near the center, and in many cases the extended 
line emission can be strongly perturbed by nongravitational forces.  For these 
objects we must rely on more indirect methods to estimate the central masses.
A promising technique employs ``reverberation mapping'' (Blandford \& McKee 
1982) to determine the size of the broad-line region, $r$, which when 
combined with a characteristic velocity dispersion of the line-emitting 
gas, $\upsilon$, as reflected in the observed line widths, yields an estimate 
of the virial mass, $M_{\rm vir}\,=\,r\upsilon^2/G$.  Choosing 
$\upsilon\,=\, {\sqrt{3}\over{2}}$FWHM(H\bet) for random, isotropic orbits, 
several studies have computed virial masses in this fashion for Seyfert~1 
nuclei (Ho 1999a; Wandel, Peterson, \& Malkan 1999) and low-redshift quasars 
(Kaspi et al. 2000).  Ho (1999a) used FWHM(H\bet) from single-epoch
spectra, whereas Wandel et al. (1999) argue that the variable component (rms) 
of the spectrum should yield a more faithful representation of the velocity 
field associated with the time lag used to calculate $r$.  As Wandel et al. 
note, however, the simple expectation that time-averaged spectra should yield 
narrower line widths than rms spectra appear not to hold in all objects.  In 
practice, it seems difficult to justify one choice over the other (Kaspi et 
al. 2000).  A more serious uncertainty lies in the choice of $\upsilon$.  
McLure \& Dunlop (2001) advocate that a disk component with 
$\upsilon\,=\,{3\over2}$FWHM(H\bet) yields \mbh\ values for AGNs that agree 
better with the \mbh--$L_{\rm bulge}$ relation for weakly active galaxies, 
thereby obviating the apparent discrepancy found by Ho (1999a) and Wandel 
(1999).  The situation is far from clear, however, since AGN masses computed 
assuming random orbits agree, to first order, with the \mbh--$\sigma_e$ 
relation (Gebhardt et al. 2000c; Nelson 2000; Ferrarese et al. 2001).  
Notwithstanding this and other possible systematic uncertainties (Krolik 2001), 
reverberation mapping appears capable of delivering \mbh\ for AGNs with an 
accuracy of a factor of $\sim2-3$.

Table~1 lists \mbh\ based on reverberation mapping for 20 Seyfert~1 galaxies 
and 17 low-redshift quasars.  For the sake of uniformity, we use 
the compilation of Kaspi et al. (2000) for the majority of the objects; a 
few not included in that study were taken from Ho (1999a).

Galaxy distances enter linearly into the mass determination for all methods 
except reverberation mapping, wherein the size scale depends only on the 
light-travel time between the central continuum source and the line-emitting 
gas.  Thus, it is important to pay close attention to the reliability of 
the adopted distances.  Whenever possible we use Tonry et al.'s (2001) 
homogeneous database of distances based on $I$-band surface brightness 
fluctuations.

\vskip 0.3cm
\subsection{Nuclear Luminosities}

Our subsequent analysis primarily will examine the connection between \mbh\ 
and two measures of the radio output, namely the absolute spectral power and 
the radio-to-optical luminosity ratio.  Since we are interested in quantities 
pertaining to the nucleus, we take great effort to assemble {\it nuclear}\ 
radio and optical data.   As emphasized in several recent studies of 
low-luminosity galactic nuclei (Ho 1999c; Ho et al. 2000, 2001; Ho \& Peng 
2001), high angular resolution is of paramount importance for isolating the 
central emission from the surrounding galaxy.  

For the radio band, we make use of interferometric data obtained at 6~cm with 
beam sizes \lax 5\asec.  The nuclear 6~cm spectral power, $P_{{\rm 6,nuc}}$, 
computed from the observed flux densities assuming isotropic emission, 
represents the integrated emission from all components considered associated 
with the ``active'' (nonstellar) nucleus.  In the case of the brighter AGNs, 
this often includes some extended structures, such as jet-like linear 
features, in addition to the central core.   For comparison with recent 
results from the literature, we have also collected radio data for the 
integrated emission from the whole galaxy (host plus nucleus), 
$P_{{\rm 6,tot}}$, which we approximate with low-resolution (beam \gax 1\amin) 
measurements.  A minority of the data were acquired at wavelengths other 
than 6~cm, and these were extrapolated to 6~cm assuming 
$f_{\nu}\,\propto\,\nu^{-0.5}$, the median spectrum between 6 and 20~cm 
for Seyfert nuclei found by Ho \& Ulvestad (2001).

Ho \& Peng (2001) demonstrated the utility of \hst\ images 

\vskip 0.3cm
\begin{figure*}[t]
\centerline{\psfig{file=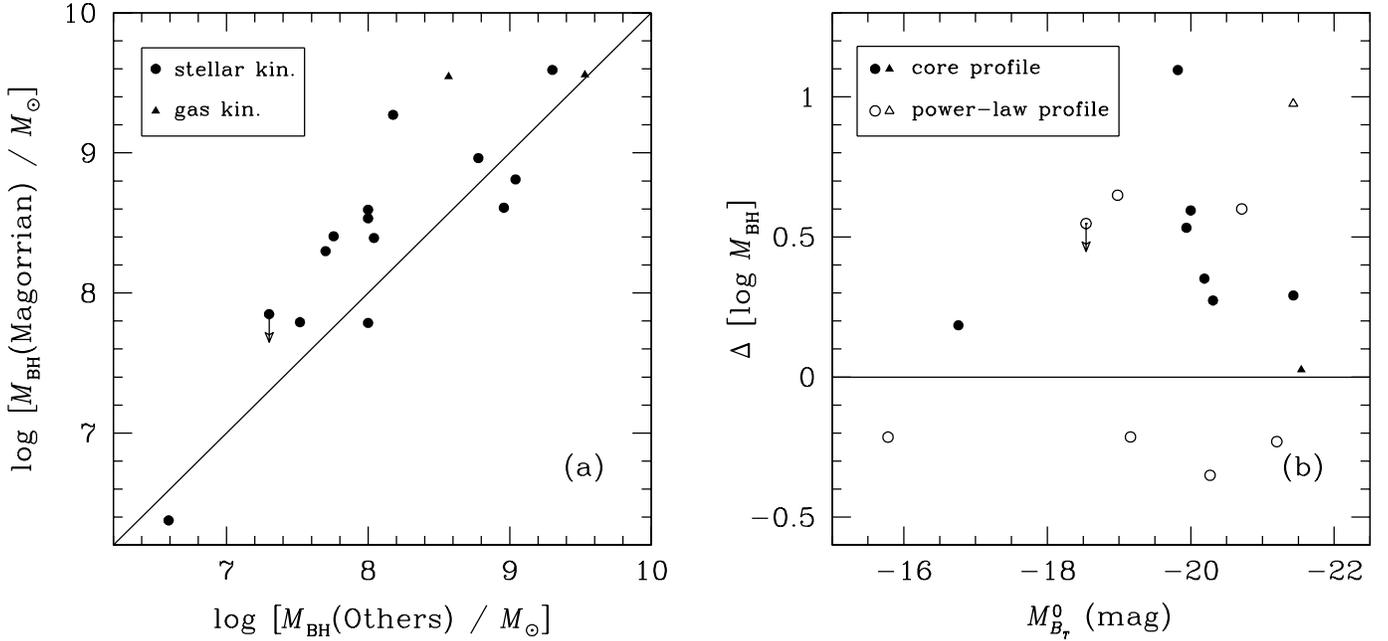,width=19.5cm,angle=270}}
\figcaption[fig1.ps]{
({\it a}) Comparison between black hole masses from Magorrian et al. (1998)
with those obtained from others.   The masses from Magorrian et al.  are
systematically higher.  ({\it b}) The difference in log~\mbh\ as a function
of the luminosity and surface brightness profile of the galaxy.  The
{\it solid line}\ denotes equality.  The most discrepant masses are found
in luminous galaxies with core-type profiles.
\label{fig1}}
\end{figure*}
\vskip 0.3cm

\noindent
for separating the
optical continuum cores in a sample of nearby Seyfert~1 galaxies.  Although 
\hst\ images are available for most of the weakly active galaxies, here we 
take a different approach.  A number of the objects have type~2 
nuclei\footnote{Type~1 and type~2 AGNs are defined as those with and without
detectable broad emission lines, respectively.}, which, according to the AGN 
unification picture (Antonucci 1993), implies that the nuclear continuum should 
be hidden from direct view, or at least appreciably extincted.  Instead, we 
estimate the continuum strength indirectly through the known (Yee 1980; Shuder 
1981) correlation between H\bet\ luminosity and $B$-band absolute 
magnitude for type~1 AGNs, as recently calibrated by Ho \& Peng (2001).  The 
line luminosity (broad and narrow components combined) is, in principle, a 
more isotropic quantity than the optical continuum luminosity.  We collected 
nuclear Balmer-line luminosities (or upper limits thereof for sources lacking 
line emission), translated them to H\bet\ if necessary\footnote{We adopt an 
intrinsic ratio of H\al/H\bet\ = 3.1, as might be appropriate for the physical 
conditions in active nuclei (e.g., Gaskell \& Ferland 1984).} (only for a 
few cases; see Table 2), and then applied the 
$L_{{\rm H}\beta}-M_B$ conversion (Ho \& Peng 2001) to arrive at 
$M^{\prime}_B$.  We use the notation $M^{\prime}_B$ to distinguish it from 
the directly measured quantity $M_B$.  Note that $M_B$ is available for all 
the quasars (Schmidt \& Green 1983) and for a number of the Seyfert~1 nuclei 
(Ho \& Peng 2001); for the bright, variable Seyfert nuclei studied 
with reverberation mapping, it can be ascertained quite reliably from the
published spectrophotometry.  For internal consistency, however, we follow the 
same procedure as adopted for the weakly active sample.

As discussed in Ho \& Peng (2001), the individual line luminosities can be 
quite uncertain, and the low-luminosity end of the $L_{{\rm H}\beta}-M_B$ 
relation has significant scatter.  We do not expect the continuum magnitudes 
to be very accurate for any given object, but it is hoped that the statistical 
results are more robust.

We use the spectral radio luminosity and the optical nuclear luminosity to 
calculate the equivalent of the standard radio-loudness parameter $R$, 
$R^{\prime} \,\equiv\,L_{\nu}({\rm 6~cm})/L^{\prime}_{\nu}(B)$.  Following 
common practice (Visnovsky et al. 1992; Stocke et al. 1992; Kellermann 
et al. 1994), we set the boundary between ``radio-loud'' and ``radio-quiet'' 
classes at $R^{\prime}$ = 10. 

\subsection{The Special Case of the Galactic Center}

The Galactic Center warrants some individual attention.  Because of its 
proximity, the ``nucleus'' of the Milky Way --- identified with Sgr~A$^*$ --- 
is overresolved with respect to other galaxies, thus rendering comparisons 
somewhat ambiguous.  In our subsequent discussions, we relax the definition 
of the Galactic Center to include increasingly larger areas surrounding 
Sgr~A$^*$.  At radio wavelengths, two distinct regions can be identified in the 
vicinity of Sgr~A$^*$, namely Sgr~A~West and Sgr~A~East (Ekers et al. 1983).  
Each of the latter two components is $\sim$40 times brighter than Sgr~A$^*$ at 
6~cm.  To achieve a linear scale roughly equivalent to that sampled in most 
of the external galaxies, however, one must extend to dimensions of 
$\sim$1\deg $\times$ 1\deg, or approximately 150~pc$\times$150~pc.  The 6~cm 
flux density on this scale (600~Jy; Mezger \& Pauls 1979) is roughly 10 times 
higher than the entire Sgr~A complex combined.

Evaluating the radio-loudness parameter for the Galactic Center requires 
knowledge of its intrinsic optical continuum luminosity, ideally measured on 
the various scales discussed above for the radio emission.  The optical 
emission, of course, is not directly observable because of the tremendous 
opacity along our line of sight.  Instead, we proceed as follows.  We utilize 
the luminosity measured in the hard X-ray ($2-10$~keV) band, corrected for 
photoelectric absorption, to estimate the intrinsic (unextincted) H\al\ 
luminosity using the relation between $L_{\rm X}(2-10$~keV) and 
$L_{{\rm H}\alpha}$ proposed by Ho et al. (2001) for nearby galactic nuclei.  
Next, $L_{{\rm H}\beta}$ follows straightforwardly from 
$L_{{\rm H}\alpha}$ 

\vskip 0.3cm
\begin{figure*}[t]
\centerline{\psfig{file=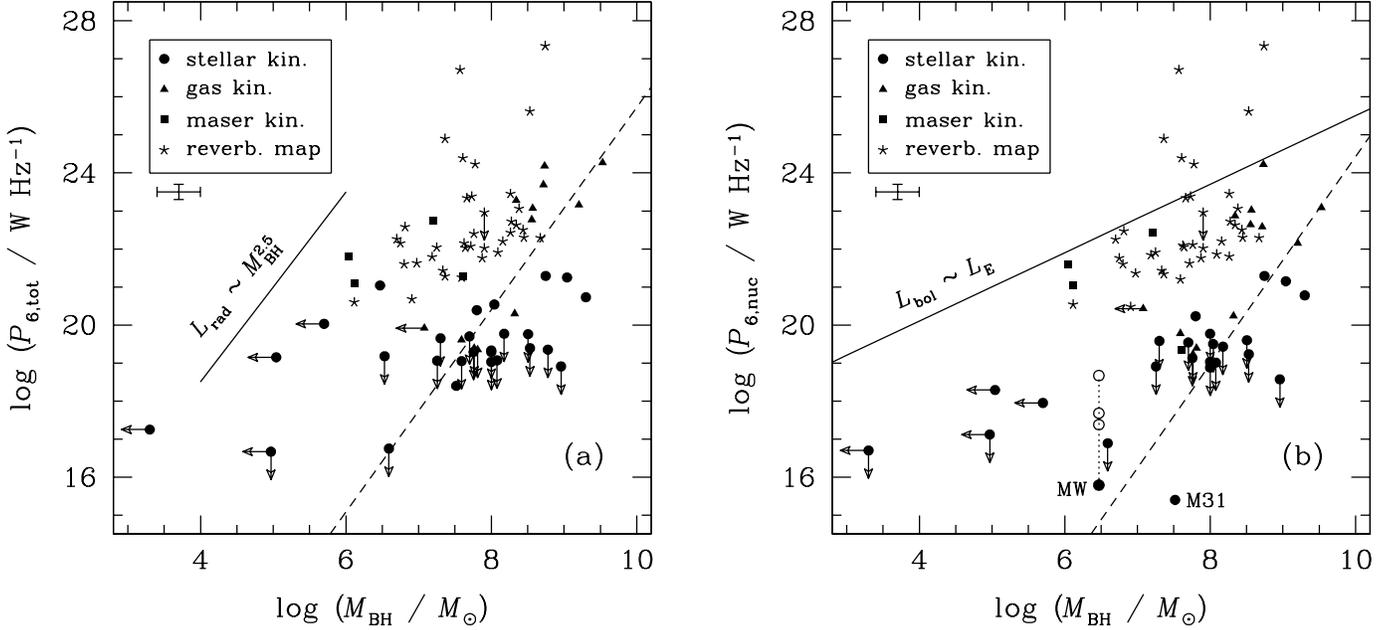,width=19.5cm,angle=270}}
\figcaption[fig2.ps]{
Dependence of ({\it a}) total and ({\it b}) nuclear 6~cm spectral
power on black hole mass.  The symbols for the different methods of mass
determination are given in the legend, below which representative error bars
are plotted.  Arrows indicate upper limits.  The {\it dashed line}\
in each graph gives the regression fit proposed by Franceschini et al. (1998);
the {\it solid lines}\ are discussed in the text.  In panel ({\it b}) we
have labeled the outlier M31.  Four values of $P_{\rm 6,nuc}$ are given
for the Milky Way (MW), connected by the {\it dotted line}; in increasing
value, they are for Sgr~A$^*$, Sgr~A~West, Sgr~A~West+East, and the central
1\deg $\times$ 1\deg\ ($\sim$150~pc$\times$150~pc) (see \S~2.3).
\label{fig2}}
\end{figure*}
\vskip 0.3cm

\noindent
for an assumed H\al/H\bet\ ratio. And finally, the
relation between $L_{{\rm H}\beta}$ and $M_B$ (Ho \& Peng 2001) yields the
$B$-band continuum flux density needed to calculate the $R$ parameter.  

Baganoff et al. (2001) recently detected Sgr~A$^*$ unambiguously in the hard 
X-rays with {\it Chandra}; the unabsorbed $2-10$~keV luminosity has a 
surprisingly low value of 2.4\e{33} \lum. The hard X-ray emission on larger 
scales comes from observations performed using {\it ASCA} by Koyama et al. 
(1996).  On dimensions which encompass Sgr~A~West, $L_{\rm X}(2-10$~keV) = 
7.7\e{35} \lum, while for the 1\deg $\times$ 1\deg\ region they find 
$L_{\rm X}(2-10$~keV) = 7.7\e{36} \lum.  We could not locate a definitive 
value for the X-ray luminosity of Sgr~A~East.  Following the above procedure, 
we find that Sgr~A$^*$, Sgr~A~West, and the 1\deg $\times$ 1\deg\ region have 
$\log R$ = 4.9, 3.8, and 4.0, respectively.  The Galactic Center, irrespective 
of one's exact definition of its boundaries, evidently is extremely 
``radio loud'' according to the conventional $R$ parameterization.

Admittedly, the above conclusions for the Galactic Center depend strongly on 
the applicability of the conversion factors used to translate 
$L_{\rm X}(2-10$~keV) to $M_B$, ones which were originally derived for more 
luminous, more active type~1 nuclei (see Ho et al. 2001; Ho \& Peng 2001).  
The spectral classification of the Galactic Center is unknown.  Ho et al. 
(2001) find that type~2 nuclei with detectable X-ray cores generally have 
$L_{\rm X}$/$L_{{\rm H}\alpha}$ values that are a factor of $\sim$10 lower 
than in type~1 objects.  Even a factor of 10 error, or greater, however, cannot 
erase the large $R$ values given above.  And although a direct relation 
between $L_{{\rm H}\beta}$ and $M_B$ does not exist for type~2 objects 
(the optical continuum is weak or obscured), to the extent that the 
line emission is powered by photoionization and $M_B$ traces the low-energy 
tail of the ionizing continuum, the two quantities should roughly scale with 
one another in type~2 objects as they do in type~1 systems.

\vskip 1.5cm
\subsection{Error Estimates}

Systematic errors affect many of the quantities used in this paper.  Although 
formal uncertainties are not specified explicitly for the data presented in 
Tables 1 and 2, we wish to alert the reader of their likely magnitude and 
impact on our analysis.  The representative error bars shown in the figures of 
this paper are meant to capture the assessment given in this section.

Ho \& Peng (2001; see \S~3.3) give a fairly thorough account of the error 
budget associated with the radio and optical luminosities.  We will not 
repeat the details here, except to reiterate that the typical uncertainties 
for the radio powers, optical line luminosities, and $R^{\prime}$ are 
$\sim$0.2, 0.3, and 0.5~dex, respectively.

At the moment, the uncertainties on \mbh\ are still quite varied.  The 
masses for the BHs in the Galactic Center (Genzel et al. 2000) and NGC 4258 
(Miyoshi et al. 1995) are known with high confidence, on the order of 10\%.
Masses based on three-integral modeling of stellar kinematics are accurate to 
$\sim$0.3~dex on average (Gebhardt et al. 2000b), but the few that are still 
based on two-integral models may undergo more significant revisions in the 
future.  While it is generally thought that gas-kinematical methods are less 
prone to modeling uncertainties than those based on stellar kinematics, under 
some circumstances our inability to treat realistically certain effects such as 
asymmetric drift may lead to serious systematic errors.  In the case of 
IC~1459, the effect of asymmetric drift on \mbh\ may be as large as a factor of 
4 (Verdoes Kleijn et al. 2000).   This correction, on the other hand, is much 
less significant in NGC~3245, whose \mbh\ has been determined to an accuracy of 
$\sim$25\% (Barth et al. 2001).  Lastly, we concluded in \S~3.1 that 
reverberation-mapping masses are probably accurate to a factor of $2-3$.  Thus, 
an overall uncertainty of 0.3~dex for \mbh\ seems

\vskip 0.3cm
\psfig{file=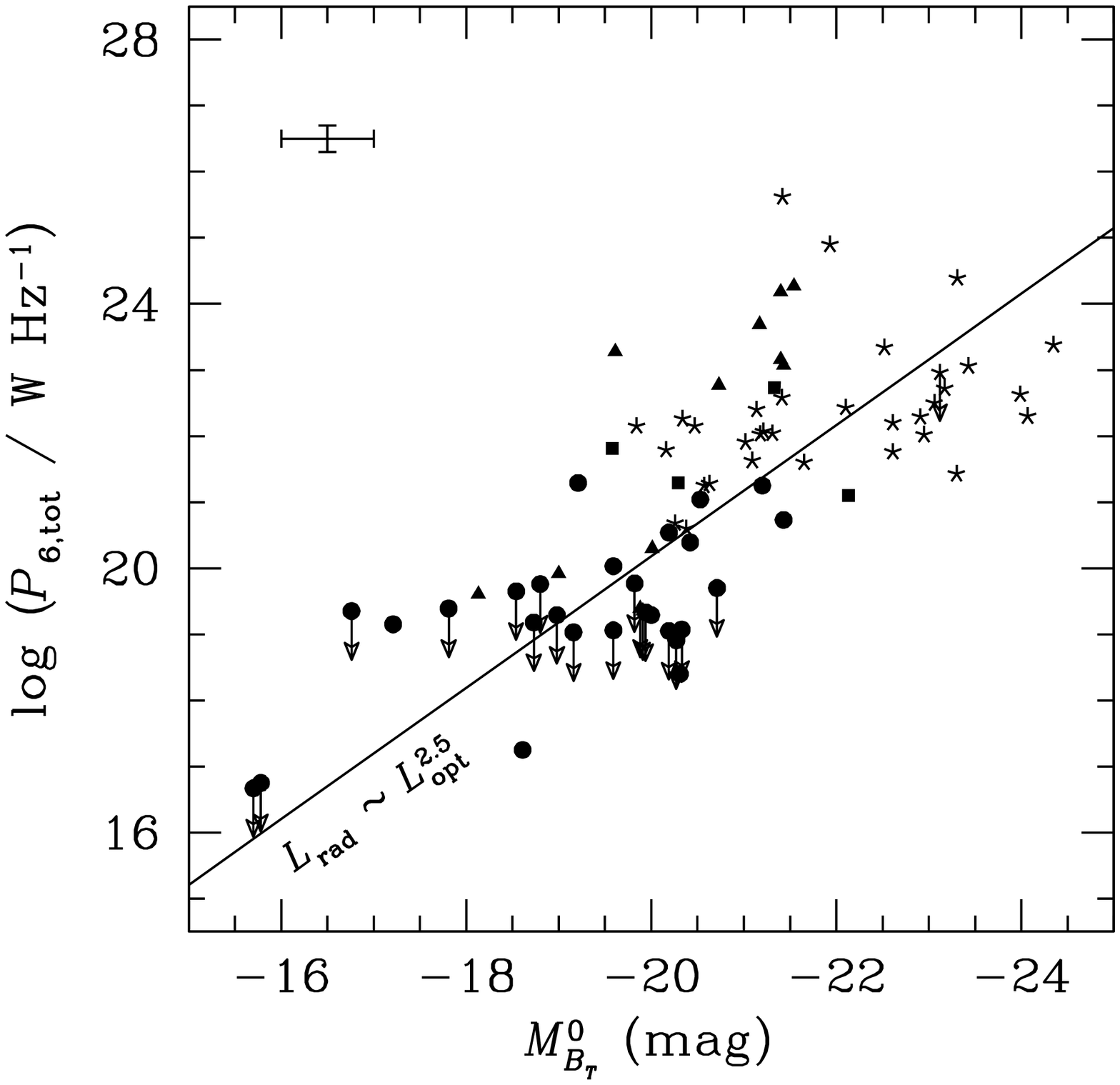,width=8.5cm,angle=0}
\figcaption[fig3.ps]{
Dependence of total 6~cm spectral power on the absolute $B$ magnitude of
the galaxy.  Symbols as in Figure~2.  Representative error bars are plotted
in the upper left corner; arrows indicate upper limits.  The {\it solid line}\
shows the formal linear regression fit.
\label{fig3}}
\vskip 0.3cm

\noindent
reasonable.

\section{Parameter Correlations}

\subsection{Radio Power and Black Hole Mass}

Concentrating first on the purported dependence of total radio power on \mbh,
Figure~2{\it a}\ shows that these two quantities are {\it not}\ well correlated 
for the sample of weakly active galaxies, contrary to the results of 
Franceschini et al. (1998).  Laor (2000) recently arrived at a similar 
conclusion using the less reliable (see \S~2.1) BH masses from Magorrian et 
al. (1998).  Although the best-fit regression line proposed by Franceschini et 
al.  ($L_{\rm rad}\, \propto\, M_{\rm BH}^{2.5}$; {\it dashed line}) does 
roughly follow the trend of the data, the scatter about the fit is enormous.
The distribution of points may be tracing an upper envelope.  The absence of 
points on the upper left corner of the diagram is real: luminous radio sources 
associated with low-mass BHs (which preferentially inhabit low-mass bulges)
can hardly be missed.  The lower right portion of the graph, on the other 
hand, is mostly populated by upper limits, and the remaining blank region may
reflect the observational bias against finding faint radio sources in the most 
luminous, on average more remote, galaxies which house the heftiest BHs.  

The form of the ridge-line of the upper envelope, which approximately follows 
$L_{\rm rad}\, \propto\, M_{\rm BH}^{2.0-2.5}$, itself can be explained as an 
indirect by-product of two known, more fundamental correlations.  Although the 
physical cause is not well understood, it has long been known that 
the integrated radio emission of early-type galaxies increases with their total 
optical luminosity (e.g., Auriemma et al. 1977; Fabbiano, Gioia, \& Trinchieri 
1989; Sadler, Jenkins, \& Kotanyi 1989; Calvani, Fasano, \& Franceschini 1989; 
Ho 1999b).  The total mass (Heckman 1983) or pressure (Whittle 1992b) of the 
bulge component has been suggested as a parameter that could affect the 
efficacy of generating radio emission.  In previous studies, the relation can be 

\vskip 0.3cm
\psfig{file=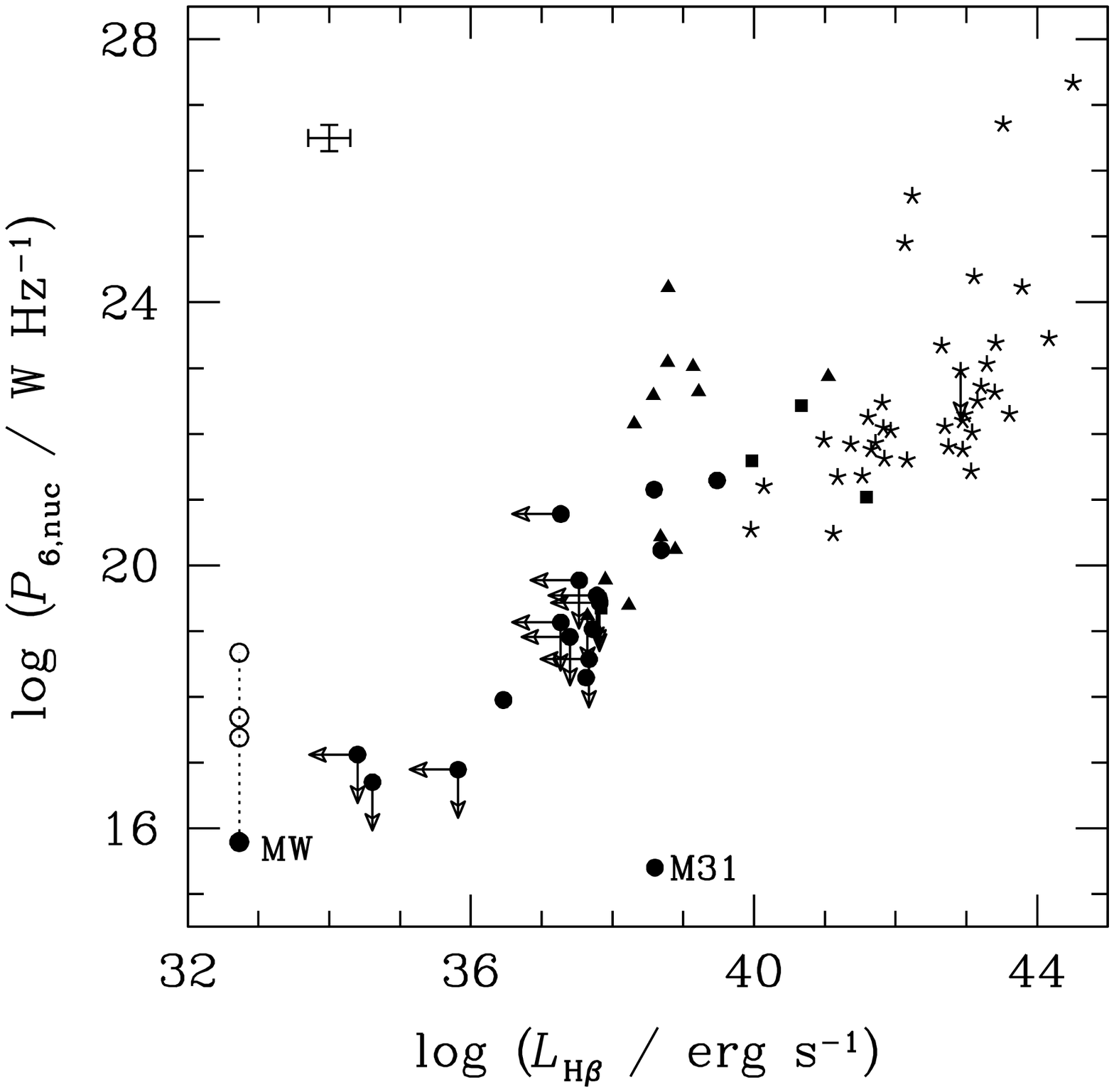,width=8.5cm,angle=0}
\figcaption[fig4.ps]{
Correlation between nuclear 6~cm spectral power and H\bet\ (broad + narrow
components) luminosity. Symbols as in Figure~2.  Representative error bars
are plotted in the upper left corner; arrows indicate upper limits.
We have labeled the outlier M31.  Four values of $P_{\rm 6,nuc}$ are given
for the Milky Way (MW), connected by the {\it dotted line}; in increasing
value, they are for Sgr~A$^*$, Sgr~A~West, Sgr~A~West+East, and the central
1\deg $\times$ 1\deg\ ($\sim$150~pc$\times$150~pc) (see \S~2.3).
\label{fig4}}
\vskip 0.3cm

\noindent 
described as $L_{\rm rad}\,\propto\,L_{\rm opt}^{\beta}$, generally with 
$\beta\,\approx\,1-3$.  Figure~3 shows the strong correlation between total 
radio emission and integrated absolute $B$ magnitude for our sample.  The 
generalized Kendall's $\tau$ test (Isobe, Feigelson, \& Nelson 1986) returns a
correlation coefficient of $-1.1$ at a significance level $>$99.99\%.
A linear regression fit ({\it solid line}) using Schmitt's (1985) method gives 
$\log P_{{\rm 6,tot}}\,\propto\,-0.99 M^0_{B_T}$, or 
$L_{\rm rad}\,\propto\,L_{\rm opt}^{2.5}$.  Now, since \mbh\ scales roughly 
linearly with the optical luminosity of the bulge (see references in \S~1), 
which in bulge-dominated systems is comparable to the integrated light of 
the galaxy, it follows that $L_{\rm rad}\,\propto\,M_{\rm BH}^{2.5}$, as 
observed (Fig.~2{\it a}; {\it solid line}).

The active galaxies in Figure~2{\it a}\ do form a fairly well defined 
correlation, but this is simply a manifestation of the fact that for these
objects the AGN component dominates the integrated emission, and, as we now 
argue, the relation between the nuclear radio power and \mbh\ {\it is}\ 
physically meaningful.  

The distribution of $P_{{\rm 6,nuc}}$ vs. \mbh\ 
also resembles an upper envelope (Fig.~2{\it b}), whose overall appearance 
is similar to that of the $P_{{\rm 6,tot}}$ vs. \mbh\ diagram.  Again, 
the weakly active galaxies do not trace the Franceschini et al. relation.  
The location of the points for Sgr~A$^*$ and the nucleus of M31 are 
particularly striking.  They are the only objects with stellar-kinematical 
masses in the range \mbh\ $\approx\, 10^6$ to few\e{7} \solmass\ that have 
detected radio cores. Both are extremely weak, with $P_{{\rm 6,nuc}}$ \lax 
$10^{16}$ W~Hz$^{-1}$, approximately 3 orders of magnitude lower than the 
upper limits for the more distant galaxies.  The position of the Galactic 
Center point, however, would migrate upwards (see points joined by dotted 
line) depending on one's definition of the ``center,'' in which case M31 would 
be a distinct outlier.  In any case, we suspect that the undetected galaxies 
could have tiny radio cores like Sgr~A$^*$ or the nucleus of M31 if they were 
to be observed at much higher sensitivity and resolution.  Although the 
distribution of points in the lower right corner of 

\vskip 0.3cm
\begin{figure*}[t]
\centerline{\psfig{file=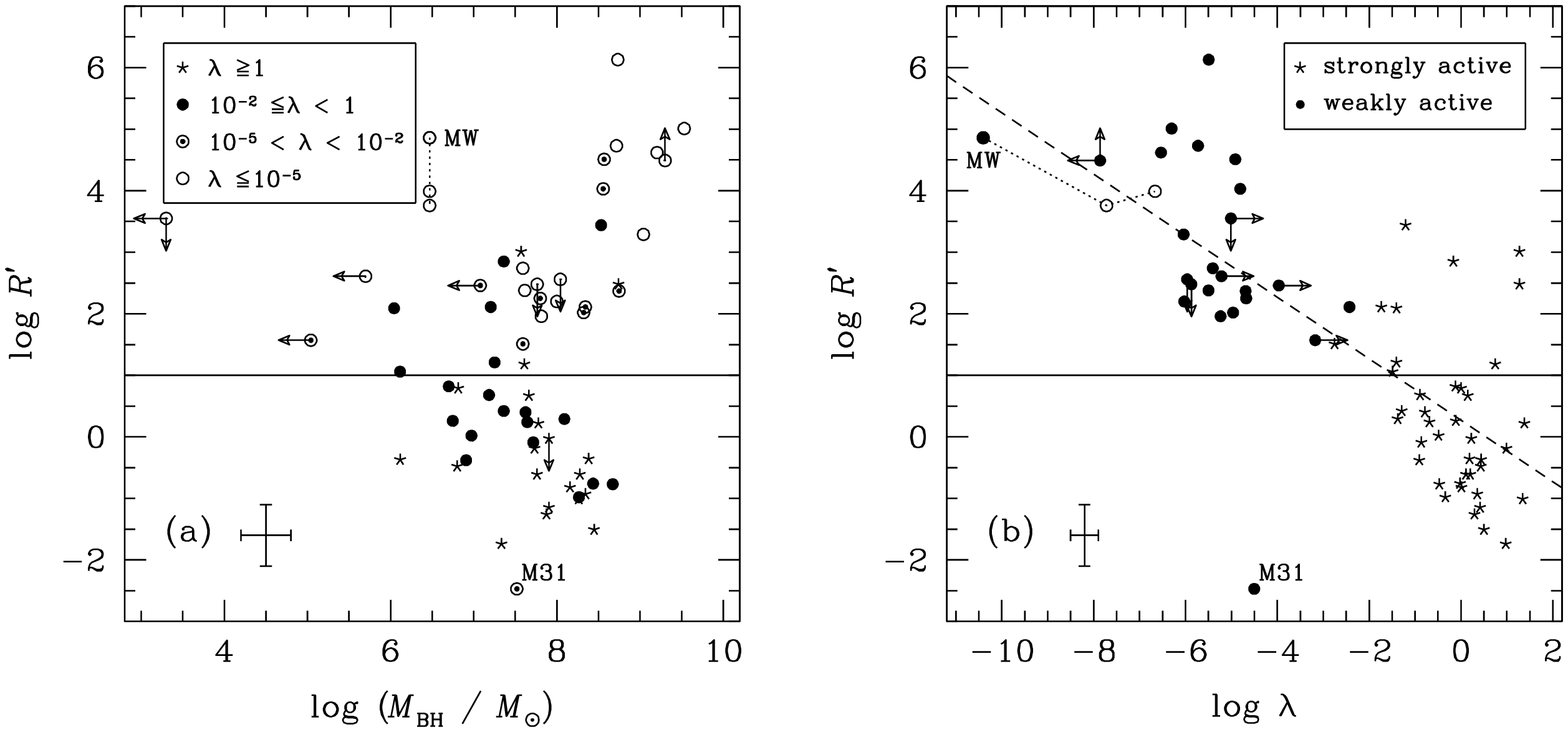,width=19.5cm,angle=270}}
\figcaption[fig5.ps]{
Distribution of the nuclear radio-to-optical luminosity ratio $R^{\prime}$
(see text) versus ({\it a}) black hole mass and ({\it b})
$\lambda\,\equiv\,L_{\rm bol}/L_{\rm E}$.  The {\it solid line}\ marks the
formal division between radio-loud and radio-quiet objects,
$R^{\prime}$ = 10.  The {\it dashed line}\ in panel ({\it b}) is the
best-fitting linear regression line.  The legends indicate the meaning of
the symbols.  Representative error bars are plotted in the lower left corner
of each panel; arrows indicate limits.  The most massive BHs tend to be
exceptionally radio-loud, highly sub-Eddington systems.  We have labeled the
outlier M31.  Three points are plotted for the Milky Way (MW), connected by
the {\it dotted line}; in increasing value of $R^{\prime}$, they are for
Sgr~A~West, the central 1\deg $\times$ 1\deg\ ($\sim$150~pc$\times$150~pc),
and Sgr~A$^*$ (see \S~2.3).
\label{fig5}}
\end{figure*}
\vskip 0.3cm

\noindent
Figure~2{\it b}\ remains 
to be determined, it is likely that \mbh\ will continue to be a very poor 
predictor of radio power.  Consider M31 and M81.  Their BH masses differ by 
only a factor of $\sim$2, but the contrast in their radio core powers is a
factor of nearly $10^5$.  
 
The ridge-line of the upper envelope, on the other hand, is fairly well
defined, and as in Figure~2{\it a}, the ``zone of avoidance'' in the upper
left corner is genuine.  Ignoring for the moment the handful of high-power
sources with $P_{{\rm 6,nuc}}\,>\,10^{24}$ W~Hz$^{-1}$, it appears that at a 
given mass, the maximum core radio power increases roughly linearly with 
\mbh.  A natural explanation for the functional form of the ridge-line can be 
found by appealing to Eddington-limited accretion and recognizing that radio 
power also scales roughly linearly with accretion (optical) luminosity for 
relatively active AGNs (Ho \& Peng 2001), or equivalently, with mass accretion 
rate.  It is worth remarking that the sample of the present study, although 
significantly more heterogeneous and diverse than that investigated by Ho 
\& Peng (2001), also displays a reasonably strong correlation between radio 
and optical luminosity (Fig.~4)\footnote{As with the results reported in 
Ho \& Peng (2001), we have confirmed that this correlation is not a spurious 
distance effect.  See Ho \& Peng for a discussion of the statistical method 
used to evaluate partial correlations with a third variable.}.  The maximum 
luminosity output of an accreting object in an isotropic, homogeneous system is 
set by the Eddington luminosity, 
$L_{\rm E} = 1.3\times10^{38} (M_{\rm BH}/M_\odot)$ \lum.  Setting 
$L_{\rm bol}\,=\,L_{\rm E}$, and approximating the bolometric luminosity with 
$L_{\rm bol}\,\approx\,c_B L_B\,\approx\,c_B c_r P_{{\rm 6,nuc}}$, 
we immediately arrive at $P_{{\rm 6,nuc}}\,\propto\,M_{\rm BH}^{1.0}$.  The 
constant $c_B$ ranges from $\sim 11-17$ for luminous AGNs and quasars (Sanders 
et al. 1989; Elvis et al. 1994) to $\sim$24 for very low-luminosity, possibly 
advection-dominated systems (median value for the 12 objects studied by Ho 
1999c and Ho et al. 2000).  The constant $c_r$ can be obtained from the 
radio-optical continuum correlation of Ho \& Peng (2001), and it differs
slightly for radio-loud compared to radio-quiet objects. The {\it solid line}\
in Figure~2{\it b}\ was obtained by choosing $c_B$ = 17 (Sanders et al. 1989) 
and the radio-loud branch of the $P_{{\rm 6,nuc}}-M_B$ relation (Ho \& Peng 
2001).  The agreement between the line and the boundary of the upper envelope 
is surprisingly good, both for the slope and the intercept.  Nearly all of the 
AGNs fall in a band bracketed by $L_{\rm bol}/L_{\rm E}\,\approx\,0.01-1$.  
Notably, many of the moderately active but lower luminosity AGNs (nearby 
Seyfert~2 nuclei and LINERs) whose masses have been determined by maser or 
ionized-gas kinematics are also broadly distributed among the more luminous 
AGNs.  By contrast, the optically and radio quiescent systems, which comprise 
all the objects with stellar-based masses, uniformly occupy the highly 
sub-Eddington regime of the diagram.  A small cluster of radio-luminous objects 
(e.g., 3C~120, 3C~273, 3C~351, 3C~390.3) lie above the 
$L_{\rm bol}\,=\,L_{\rm E}$ line, but this is not unexpected because objects 
with powerful radio jets are known to follow a steeper radio-optical 
correlation (e.g., Serjeant et al. 1998; Willott et al. 1999).

\subsection{Radio Loudness and Black Hole Mass}

BH accretion in galactic nuclei invariably generates radio emission.  The 
fundamental parameters responsible for the tremendous range of the observed 
strength of the radio output, however, are not well established and have been 
largely a subject of speculation.  Recent advances in high-resolution imaging 
of quasars consistently suggest that radio-loud objects reside in hosts which 
lie on the top end of the galaxy luminosity function, whereas the hosts of 
radio-quiet sources generally span a wider range of luminosities.  The degree 
of radio loudness, therefore, seems to depend on galaxy mass.  In view of the 
link between BH mass and bulge mass (see \S~1), it is reasonable to deduce 
that radio loudness would depend on \mbh.  Laor (2000) examined this issue 
using a sample of low-redshift ($z < 0.5$) quasars from the Palomar-Green (PG)
survey (Schmidt \& Green 1983).  The radio properties of the PG sources are 
known (Kellermann et al. 1989), and approximate virial masses can be estimated 
from the H\bet\ line widths and optical continuum luminosities following 
empirical calibrations derived from AGN variability studies (Laor 1998; 
Kaspi et al. 2000).  Laor (2000) concluded that the radio-loudness parameter 
$R$ is a strong function of \mbh: most radio-loud objects (defined by 
$\log R\,>\,1$) have \mbh\ $>\,10^9$ \solmass, whereas nearly all quasars with 
\mbh\ $<\,3 \times 10^8$ \solmass\ are radio quiet.

The sample considered in this study, which spans a much wider gamut of activity 
level, yields a more complex picture.  As shown in Figure~5{\it a}, the clean 
segregation in mass between radio-loud and radio-quiet objects suggested by 
Laor disappears.  The radio-loud plane ($\log R^{\prime}\,>\,1$) is richly 
populated with masses ranging from log~\mbh\ $\approx$ 9.5 to 6.0, and possibly 
even lower if we consider the upper limits.  Indeed, most or all objects with 
log~\mbh\ \gax\ 8.5 are radio loud, but radio-loud objects are by no means 
restricted to the high-mass domain.  

To explore other factors which may govern the distribution of points on 
this plot, we have coded the symbols according to the bolometric luminosity of 
the nucleus normalized to the Eddington luminosity, 
$\lambda\,\equiv\,L_{\rm bol}/L_{\rm E}$.  As in \S~3.1, we estimate 
$L_{\rm bol}$ crudely from $L_B$ (which is based on $L_{{\rm H}\beta}$).  An 
intriguing pattern emerges.  The majority of sources with \lamb\ \gax\ 1 land 
in the radio-quiet regime; those with \lamb\ \lax\ 10$^{-5}$ fall exclusively 
in radio-loud territory; and objects with intermediate values of \lamb\ 
straddle the (somewhat arbitrary) $R^{\prime}$ boundary.  (M31 is a persistent 
outlier.)  The dependence of $R^{\prime}$ on \lamb\ is shown explictly in 
Figure~5{\it b}; the objects form a striking inverse correlation, albeit with 
substantial scatter. The majority of the strongly active nuclei (Seyfert~1s 
and quasars) are characterized by log~\lamb\ \gax\ $-2$ and 
$\log R^{\prime}\,<\,1$.  With the exception of the deviant point for M31, 
{\it all}\ of the weakly active objects (those with \mbh\ based on 
spatially resolved kinematics) are confined to log~\lamb\ $<\,-2$ and 
$\log R^{\prime}\,>\,1$.  More quantitatively, the generalized Kendall's 
$\tau$ test, with M31 omitted, gives a correlation coefficient of $-0.97$ at a 
significance level $>$99.99\%.  A linear regression fit ({\it dashed line}) 
using Schmitt's (1985) method, which treats censoring in both variables, gives 

$$\log R^{\prime}\,=\, -(0.50\pm0.07) \log \lambda\,+\, (0.27\pm0.24).$$

\noindent
Since \lamb\ depends on the mass accretion rate $\dot{M}$ (see, e.g., Fig.~7 
of Narayan et al. 1998b), an immediate consequence of the $R^{\prime}-\lambda$ 
inverse correlation is that the degree of radio loudness depends strongly
on $\dot{M}$.

Although the strong inverse correlation between $R^{\prime}$ and $\lambda$ may 
superficially resemble a mutual dependence of these variables with optical 
luminosity ($R^{\prime} \propto L_B^{-1}$, $\lambda \propto L_B$), we note that 
the slope of the $R^{\prime}-\lambda$ relation differs significantly from $-1$.
We do not believe that this is the primary driver of the observed correlation.


\section{Discussion and Summary}

A number of statistical studies of luminous AGNs indicate that the processes 
responsible for the generation of radio sources depend directly on BH mass 
(McLure et al. 1999; Nelson 2000; Laor 2000; McLure \& Dunlop 2001).  Lacy et 
al. (2001) suggest that, in addition to the primary dependence on \mbh, the 
radio luminosity is also a weak function of $L/L_{\rm E}$.   Along similar 
lines, Franceschini et al. (1998) proposed that the weakly active nuclei in 
nearby galaxies also obey a correlation between radio luminosity and \mbh, 
roughly of the form $L_{\rm rad}\, \propto\, M_{\rm BH}^{2.5}$.  The 
correlation evidently holds for the core radio emission as well as for the 
integrated radio emission.  Franceschini et al. (1998) and Di~Matteo et al. 
(2001) argue that the slope of the $L_{\rm rad}-M_{\rm BH}$ relation can be 
explained qualitatively in the context of advection-dominated accretion of hot 
plasma undergoing Bondi-type inflow.  This interpretation, however, has been 
questioned by Yi \& Boughn (1999), who noted that, given the observed radio 
powers and the critical assumption of Bondi accretion, the ADAF model predicts
far more X-ray radiation than is actually detected.   

We have reexamined these issues using a comprehensive compilation of 
up-to-date BH masses and photometric parameters.  Our sample includes all 
the nearby, weakly active or inactive galaxies which have reliable BH 
masses determined through spatially resolved kinematics, as well as all the 
AGNs (Seyfert~1 nuclei and quasars) for which virial masses have been 
derived through reverberation mapping.  Our main results lead to the 
following conclusions.

1.  There is no simple relation between integrated radio luminosity and \mbh.  
The distribution of objects is consistent with either an upper envelope or a 
loose correlation of the form $L_{\rm rad}\, \propto\, M_{\rm BH}^{2.0-2.5}$,
similar to that found by Franceschini et al. (1998), but we offer a different 
interpretation.  We suggest that the integrated $L_{\rm rad}-M_{\rm BH}$ 
relation arises indirectly from two known, physically more fundamental 
correlations, namely that between integrated radio luminosity and optical bulge
luminosity (mass) and that between bulge luminosity (mass) and BH mass.

2.  There is no simple relation between core radio luminosity and \mbh.  
The distribution of objects appears to follow an upper envelope defined 
by Eddington-limited accretion.  At any given value of \mbh, the maximum 
luminosity attained is set by $L_{\rm bol}\,=\,L_{\rm E}$, and as 
discussed by Ho \& Peng (2001), the core radio power traces the accretion 
luminosity.  The majority of the more active nuclei in our sample appear to 
be characterized by $L_{\rm bol}\,\approx\,(0.01-1) L_{\rm E}$.  Not 
surprisingly, the more quiescent objects are radiating at only a tiny fraction 
of the Eddington rate.

3.  There is no simple relation between the radio-loudness parameter $R$ 
and \mbh. Specifically, we do not find the clean division between $R$ and 
\mbh\ suggested by Laor (2000).  Radio-loud nuclei are not confined solely 
to galaxies with the most massive BHs, but instead can inhabit galaxies with 
\mbh\ as low as $10^6$ \solmass, or perhaps even less.  The nucleus of the 
Milky Way provides an interesting, if unfamiliar, illustration.   As a 
corollary of this result, radio-loud nuclei are not restricted to early-type 
(S0 and elliptical) galaxies, as conventionally thought; rather, they can 
be hosted by galaxies of a variety of morphological types, including 
disk-dominated spirals.  Both of these results have been foreshadowed by 
recent investigations of the nuclear spectral energy distributions of nearby 
low-luminosity AGNs (Ho 1999c; Ho et al. 2000).  Ho \& Peng (2001) 
specifically challenged the traditional notion that Seyfert nuclei, the 
majority of which are hosted by disk galaxies,  are primarily radio-quiet 
objects.  Using high-resolution optical and radio measurements of a 
well-defined set of Seyfert~1 galaxies, they showed that the nuclear $R$ 
parameter places more than 60\% of the objects in the category of radio-loud 
sources ($\log R > 1$).  The present study proceeds in a similar spirit.  
In order to assess the relative radio power, we constructed nuclear measurements 
of $R^{\prime}$, which differs from $R$ only in that the optical continuum 
luminosity was obtained indirectly through the H\bet\ luminosity.  When 
studying nearby galaxies with low-luminosity nuclei, we reiterate the 
importance of using high-resolution data to properly define the {\it nuclear}\ 
$R$ parameter --- this is the quantity that is relevant for comparison with 
luminous AGNs and quasars, whose nonstellar nuclei dominate the integrated 
emission.  We disagree with Laor's (2000) use of single-dish radio measurements 
to evaluate the radio loudness of the objects in the sample of Ho (1999c).  
Laor also questioned the utility of the standard $R$ parameter as applied to 
low-luminosity AGNs because he suspected that the optical bolometric 
correction may be exceptionally large for these objects.  The 
characteristic weakness of the ``big blue bump'' in the spectral 
energy distributions of low-luminosity AGNs (Ho 1999c, 2001; Ho et al. 2000) 
indeed does lead to a larger value of $L_{\rm bol}/L_B$ than is typically 
seen in higher luminosity sources, but this difference is only a factor of 
$\sim$2 (see \S~3.1), which is insufficient to alter the main conclusions.  

4. We find a striking inverse correlation between $R^{\prime}$ and 
$\lambda\,\equiv\,L_{\rm bol}/L_{\rm E}$.  Since \lamb\ varies as a function 
of the mass accretion rate $\dot{M}$, the most straightforward implication of 
this result is that the relative radio power increases with decreasing $\dot{M}$.
A significant fraction of the strongly active sources have high accretion 
rates (log~\lamb\ \gax\ $-2$) and are radio quiet ($\log R^{\prime}\,<\,1$), 
whereas nearly all of the weakly active objects are starved for fuel (all 
log~\lamb\ $<\,-2$) and are radio loud ($\log R^{\prime}\,>\,1$).  

5. The systematic dependence of $R$ on \lamb\ and the tendency for local 
galaxies to have nuclei which are both underluminous {\it and}\ radio 
loud are qualitatively consistent with predictions from accretion-disk 
theory.  The amount of fuel available in the centers of present-day 
galaxies is plausibly quite low.  As the accretion rate falls below a critical 
threshold of $\dot{M}\,\approx\,0.1 \alpha^2 \dot{M_{\rm E}}\,\approx\,0.01 
\dot{M_{\rm E}}$, where $\alpha$ is the standard viscosity parameter (assumed 
to have a value $\sim$0.3), the accretion flow makes a transition to an 
optically thin, two-temperature ADAF (Narayan et al. 1998b).  Under these 
conditions, the low density and low optical depth of the accreting material 
lead to inefficient cooling, the radiative efficiency is much less than the 
canonical value of 10\%, and thus the resulting luminosity is low.  Moreover, 
ADAFs produce generically ``radio-loud'' spectral energy distributions,
for two reasons.  First, cyclo-synchrotron emission at radio wavelengths 
provides energetically important cooling.  And second, the broad-band 
spectrum, by definition, lacks the big blue bump usually attributed to thermal 
emission from the optically thick, geometrically thin disk (Shields 1978; 
Malkan \& Sargent 1982).  ADAFs are naturally bright in the radio and dim at 
optical/ultraviolet wavelengths: both conspire to boost $R$.  The 
optical/ultraviolet component from an ADAF comes from inverse Compton 
scattering of the cyclo-synchrotron photons, and its strength increases 
sensitively with rising $\dot{M}/\dot{M}_{\rm Edd}$ (see e.g., Fig.~1 of 
Mahadevan 1997). This qualitatively explains the inverse correlation between 
$R$ and \lamb.  Furthermore, Rees et al. (1982) have suggested that the 
vertically thick structure of the ``ion torus'' may help facilitate the 
collimation of radio jets.  Finally, we note that the majority of the weakly 
active objects in our sample have values of \lamb\ \lax\ $10^{-4}-10^{-3}$, 
comfortably below the threshold within which ADAFs operate. Although the above 
general arguments need to be confirmed with more quantitative calculations, a 
number of authors have invoked ADAF models to fit the spectral energy 
distributions of some of the objects in our sample (Sgr~A$^*$, Manmoto, 
Mineshige, \& Kusunose 1997, Narayan et al. 1998a; NGC 4258, Lasota et al. 
1996, Chary et al. 2000; M81, Quataert et al. 1999; M87 and NGC 4649, 
Di~Matteo et al. 2000).

\acknowledgements
I thank Marianne Vestergaard for discussions which motivated me to examine the 
issues discussed in this paper.  Gary Bower kindly communicated information on 
M81 and NGC 3998 in advance of publication, and Swara Ravindranath helped to 
obtain \hst\ photometry for a few objects.  My research is supported in part by 
NASA grants HST-GO-06837.04-A, HST-AR-07527.03-A, and HST-AR-08361.02-A, 
awarded by the Space Telescope Science Institute, which is operated by AURA, 
Inc., under NASA contract NAS5-26555.  This work made use of the NASA/IPAC 
Extragalactic Database (NED) which is operated by the Jet Propulsion 
Laboratory, California Institute of Technology, under contract with NASA, and 
of the Lyon-Meudon Extragalactic Database (LEDA).

\end{document}